\documentclass[twocolumn,showpacs,preprintnumbers,amsmath,amssymb]{revtex4}
\usepackage{graphicx}

\begin{document}

\title{Manipulating the critical temperature for the superfluid phase transition
in trapped atomic Fermi gases}
\author{C. P. Search, H. Pu, W. Zhang, B. P. Anderson, and P. Meystre}
\affiliation{Optical Sciences Center, The University of Arizona,
Tucson, AZ 85721}
\date{\today}

\begin{abstract}
We examine the effect of the trapping potential on the critical
temperature, $T_C$, for the BCS transition to a superfluid state
in trapped atomic gases of fermions. $T_C$ for an arbitrary power
law trap is calculated in the Thomas-Fermi approximation. For
anharmonic traps, $T_C$ can be increased by several orders of
magnitude in comparison to a harmonic trap. Our theoretical
results indicate that, in practice, one could manipulate the
critical temperature for the BCS phase transition by shaping the
traps confining the atomic Fermi gases.
\end{abstract}
 \maketitle

Since the observation of Bose-Einstein condensation in trapped
atomic gases \cite{BEC}, there has been increasing interest in the
possibility of observing the Bardeen-Cooper-Schrieffer (BCS) phase
transition to the superfluid state \cite{bardeen} in dilute
fermionic akali gases
\cite{Li6,burnett,you,baranov,efremov,heiselberg,bohn,holland,mackie}.
Currently, experimental efforts in cooling fermionic atoms of
$^{6}$Li \cite{truscott,thomas} and $^{40}$K \cite{jin,jin2} to
the quantum degenerate regime have made significant progress,
reaching temperatures as low as $0.2T_{F}$ where $T_{F}$ is the
Fermi temperature. However, these temperatures are still far
beyond the critical temperature required for the BCS phase
transition, which is {\em at least} an order of magnitude lower
\cite{Li6,burnett,you,baranov,efremov}.

On the other hand, the current techniques used to cool fermionic
atoms lose their utility when the temperature becomes much less
than $T_F$. For example, evaporative cooling, which has been used
for $^{40}$K, becomes ineffective for temperatures much less than
$T_F$ due to Pauli blocking \cite{evaporate} while sympathetic
cooling of the fermions using bosons, as done with $^{6}$Li
\cite{truscott}, loses its effectiveness when the heat capacity of
the bosons falls below that of the fermions. Due to these
technical obstacles in cooling fermionic atoms, it seems that one
must seek other avenues to reach the BCS superfluid state.

Recently there have been a number of proposals to study the
possibility of achieving a higher $T_C$ by increasing the strength
of the attractive interactions, i.e., the scattering length
between the fermions using either a Feshbach resonance
\cite{bohn,holland} or photoassociation \cite{mackie}. Since $T_C$
is a function of the attractive interaction between fermions and
the density of states at the Fermi energy, $\rho_F$, higher
transition temperatures can also be achieved by increasing the
density of states for the fermions at the Fermi surface. In bulk
superconductors, the only way to modify the density of states is
by changing the spatial dimensionality of the system. However, in
atomic systems the density of states can be controlled through the
shape and strength of the external trapping potential.

The ability of an external trapping potential to change the
critical temperature for a phase transition by modifying the
density of states, $\rho(\epsilon )$ (see Eq.
(\ref{densitystates}) below), was first considered by Bagnato {\it
et al.} for Bose-Einstein condensation \cite{bagnato}. They showed
that the critical temperature for Bose-Einstein condensation,
$T_{BEC}$, as a function of the number of atoms confined in a
power law trap, $N_B$, had the power law form $T_{BEC}\sim
N_B^{\eta}$ where $2/9\leq \eta \leq 2/3$. In a similar manner, we
will show that for a trapped gas of $N_F$ fermions the density of
states at the Fermi surface has the form $\rho_F\sim N_F^{\beta}$
where $\beta$ varies from 1/9 to 1/3 depending on the type of
trap. Since $T_C$ depends {\em exponentially} on $\rho_F$, small
changes in $\beta$ can have a dramatic effect on $T_C$. In all the
experiments with trapped fermions, the trapping potential is well
approximated as being harmonic. Our results indicate, however,
that higher values of $T_C/T_F$ can be obtained using anharmonic
power law traps. This indicates that the use of anharmonic traps
could place $T_C$ within the range of current experiments.

At the ultracold temperatures achieved in fermion experiments,
$p$-wave collisions between atoms are highly suppressed and
$s$-wave collisions between atoms in the same internal state are
forbidden by the Pauli exclusion principle. Since the BCS
transition requires an attractive interaction between fermions in
order to form Cooper pairs, the most likely candidate for Cooper
pairing in alkali gases is an attractive $s$-wave interaction
between atoms in different hyperfine states. Fortunately, both
$^{6}$Li and $^{40}$K appear to be very promising candidates.
$^{6}$Li possesses an anomalously large and negative $s$-wave
scattering length equal to $-2160a_0$ \cite{Li6}, where $a_0$ is
the Bohr radius, while for $^{40}$K, a Feshbach resonance exists
for two of the hyperfine states which can be used to create the
required large attractive interaction \cite{bohn}.

Therefore, we discuss the case of s-wave pairing between atoms of
mass $m$ in two hyperfine states with an s-wave scattering length
$a<0$ for collisions between atoms in different states.
Furthermore, we assume that the number of atoms in each of the
hyperfine states, $N_F$, is the same since this is the optimal
condition for Cooper pairing. In this case, the transition
temperature for a dilute homogenous gas is given by
\begin{eqnarray}
\frac{T_C}{T_F}= \alpha\exp\left(-\frac{1}{g\rho_F}\right),
\label{TC}
\end{eqnarray}
where $g=4\pi\hbar^2|a|/m$ is the coupling constant for the
two-body interactions and $\alpha =
2^{7/3}e^{C-7/3}/\pi\approx0.28$ with $C$ being the Euler's
constant \cite{gorkov,heiselberg,lifshitz}. The density of states
at the Fermi surface, in terms of the Fermi momentum $\hbar k_F$,
is \[\rho_F=mk_F/2\pi^2\hbar^2. \] From Eq. (\ref{TC}) we can see
that by raising the density of states at the Fermi surface (or
equivalently, by raising $k_F$), there will be an exponential
increase in the transition temperature. Note that even though we
restrict ourselves to the case of $s$-wave pairing, our results
can easily be extended to pairing via higher partial waves since
in those cases the critical temperature has a similar exponential
dependence on $\rho_F$ \cite{lifshitz}.

The goal of this paper is to determine the critical temperature
for the inhomogeneous Fermi gas where the atoms in both hyperfine
states are subject to the trapping potential,
\begin{equation}
U({\bf r})=\varepsilon_1 \left|\frac{x}{a}\right|^p +
\varepsilon_2 \left|\frac{y}{b}\right|^l +\varepsilon_3
\left|\frac{z}{c}\right|^q ,\label{trap}
\end{equation}
and $p$, $l$, and $q$ are positive integers. We refer to the
magnitude of the exponents $p$, $l$, and $q$ as the confining
power of the trap. For simplicity, we assume that the trapping
potential is independent of the hyperfine state, hence the density
of fermions is the same for both states, $n({\bf r})$. If $n({\bf
r})$ varies sufficiently slowly, then we can make a local density
(or Thomas-Fermi) approximation by assuming that at each point in
space the Fermi gas can be treated as being homogenous
\cite{butts}. In that case, a local value of the critical
temperature, $T_C({\bf r})$, can be calculated using
Eq.(\ref{TC}). For temperatures $T\ll T_F$, the local density
approximation is valid provided the average distance between
atoms, $\sim k_F^{-1}$, is much less than the distances over which
$n({\bf r})$ changes significantly. Since the changes in $n({\bf
r})$ are determined by $U({\bf r})$ for which the characteristic
length scales are $a$, $b$, and $c$, we have the condition
$k_F^{-1}\ll(abc)^{1/3}$. In addition, for $T\ll T_C$, the
coherence length for the Cooper pairs, $\xi_0=\hbar^2 k_F/\pi m
\Delta_0$ where $\Delta_0=(\pi e^{-C})k_BT_C$ is the BCS gap at
$T=0$, should also be much less than $(abc)^{1/3}$
\cite{Li6,burnett}.

In the Thomas-Fermi approximation, the chemical potential, $\mu$,
is replaced with a local value, $\mu({\bf r})=\mu-U({\bf
r})+gn({\bf r})$ so that the fermions can be treated at each $\bf
r$ as an ideal homogenous gas with chemical potential $\mu({\bf
r})$. The term $gn({\bf r})$ represents the Hartree-Fock potential
experienced by each atom due to the atoms in the opposite
hyperfine state. Since we are interested in temperatures $T\ll
T_F$, the chemical potential may be approximated by the Fermi
energy, $\mu \simeq E_F=k_BT_F=\hbar^2k_F^2/2m$, since for low
temperatures $\mu=E_F+O(T/T_F)^2$ \cite{mli}. Correspondingly, we
can express $\mu({\bf r})$ in terms of a local Fermi wave number,
$k_F({\bf r})$, defined as
\begin{equation}
\frac{\hbar^2k_F^2({\bf r})}{2m}=E_F-U({\bf r})+gn({\bf r}).
\label{kfr}
\end{equation}
Using the result $n({\bf r})=k_F^3({\bf r})/(6\pi^2)$ for an ideal
Fermi gas, then gives an expression for the density
\begin{equation}
n({\bf r})\left(1-\frac{g}{A^{2/3}}n^{1/3}({\bf r})\right)^{3/2}
=\frac{1}{A} \left[E_F-U({\bf r})\right]^{3/2} ,\label{density}
\end{equation}
where $A \equiv 6\pi^2 \left( \hbar^2/2m \right)^{3/2}$. For a
dilute gas where $|a|n(0)^{1/3}\ll 1$, the Hartree-Fock term in
Eqs. (\ref{kfr}) and (\ref{density}) can be neglected in a first
approximation. The Fermi energy is then determined by the
requirement that the total number of atoms in each spin component
be conserved,
\begin{equation}
N_F=\frac{1}{A} \int_{V(E_F)} \left[E_F-U({\bf
r})\right]^{3/2}d^3r, \label{number}
\end{equation}
where the integration volume $V(E_F)$ is the volume available to a
classical particle with total energy $E_F$, i.e. $E_F\geq U({\bf
r})$. Equation (\ref{number}) can easily be solved for $E_F$ for
the case of the power law trap (\ref{trap}) and gives,
\begin{equation}
E_F=\left[\frac{A\varepsilon_1^{1/p}\varepsilon_2^{1/l}
\varepsilon_3^{1/q}\Gamma(\delta+1)}
{\Gamma(1+1/p)\Gamma(1+1/l)\Gamma(1+1/q)\Gamma(5/2)}
\left(\frac{N_F}{8abc}\right)\right]^{1/\delta} \label{EF}
\end{equation}
where $\Gamma(x)$ is the Gamma function and
\[\delta \equiv 3/2+1/p+1/l+1/q.\] Note that the limit
$p,l,q\rightarrow \infty$ corresponds to a box with volume $8abc$
\cite{bagnato}. For a harmonic oscillator, $p=l=q=2$, $8abc$ is
equal to $2^{9/2}\ell_x\ell_y\ell_z$ where
$\ell_i=\sqrt{\hbar/m\omega_i}$ is the harmonic oscillator length
along the axis of the trap with frequency $\omega_i$. Therefore,
$\bar{n}=N_F/(8abc)$ can be used to define a characteristic atomic
density. Note that, with the exception of the rigid box, $\bar{n}$
does not correspond to the average density of atoms in the trap.
In what follows, we will assume that $\bar{n}$ is fixed and
independent of $\delta$.

Equation (\ref{EF}) along with Eq. (\ref{kfr}) can be used to
calculate $T_C({\bf r})$. It is clear from Eq. (\ref{kfr}) that
$k_F({\bf r})$ is a maximum at the center of the trap and hence,
the density of states at the local Fermi surface, $\rho_F({\bf
r})=mk_F({\bf r})/2\pi^2\hbar^2$, is a maximum there.
Consequently, $T_C({\bf r})$ is largest at ${\bf r}=0$.
Physically, this means that as the temperature is lowered Cooper
pairing first occurs at the center of the trap and then spreads to
the edges of the trap as the temperature is lowered still further.
Therefore at ${\bf r}=0$ and neglecting the Hartree-Fock term in
Eq. (\ref{kfr}), the transition temperature is simply,
\begin{equation}
\frac{T_C(0)}{T_F}=\alpha\exp\left(-\frac{\hbar\pi}{|a|\sqrt{8mE_F}}\right).
\label{TCtrap}
\end{equation}
Evaluating the critical temperature at the center of the trap has
the added benefit that the Thomas-Fermi approximation is expected
to be most accurate here.

From Eqs. (\ref{EF}) and (\ref{TCtrap}) we see that for a given
value of $\bar{n}$, $T_C(0)/T_F$ is an increasing function of
$1/\delta$. Furthermore, $T_F=E_F/k_B$ is also an increasing
function of $1/\delta$. Therefore, increasing the confining power
of the trap increases not only the ratio of the critical
temperature to the Fermi temperature but also the absolute value
of the critical temperature. More importantly, from the
perspective of current experiments, anharmonic traps for which
$p,l,q\geq 3$ will result in higher values of $T_C$ (assuming that
all the other terms in Eq. (\ref{EF}) are the same). This is the
central result of this paper.

To illustrate the effect of varying the confining power of the
trap, we consider the case of $p=l=q$ for the values of 1 through
5 and $\infty$. We choose values of $\varepsilon_i$ and $a$, $b$,
and $c$ that correspond to parameters used in current experiments
with harmonic traps. In Tables~\ref{table1} and \ref{table2} we
calculate the Fermi temperature, $T_F$, as well as $T_C(0)$ for
$^6$Li and $^{40}$K, respectively. For $^6$Li we consider an
isotropic trap with $N_F=10^5$ and
$\varepsilon_i=\hbar\omega=\hbar(2\pi\times 100s^{-1})$, which
gives $a=b=c=\sqrt{2\hbar/m\omega}=5.8\mu m$ and
$\bar{n}=6.4\times 10^{13}cm^{-3}$. For $^{40}$K, we use values
from the experiment by DeMarco and Jin \cite{jin}. This gives
$N_F=3.5\times 10^5$ and
$\varepsilon_1=\varepsilon_2=\hbar\omega_r=\hbar(2\pi\times
127s^{-1})$ and $\varepsilon_3=\hbar\omega_z=\hbar(2\pi\times
19.5s^{-1})$. This gives values for the characteristic lengths of
$a=b=2\mu m$ and $c=5.09\mu m$ and a characteristic density of
$\bar{n}=2.15\times 10^{15} cm^{-3}$. Bohn has recently showed
that there exists an experimentally accessible Feshbach resonance
for two of the hyperfine states of $^{40}$K that could be accessed
to create a scattering length of $a=-1000a_0$ \cite{bohn}. We
therefore adopt this value for $a$ since in the absence of a
Feshbach resonance, the scattering lengths for $^{40}$K would
result in unreasonably small values of $T_C$.

\begin{table*}
\caption{\label{table1} $T_F$ and $T_C(0)$ as functions of the
confining power of the isotropic power-law trap, for $^6$Li.}
\begin{ruledtabular}
\begin{tabular}{ccccc}
$p=l=q$ & $E_F$(J) & $T_F$(K) & $T_C(0)/T_F$ & $T_C(0)$(K) \\
\hline 1 & $1.90 \times 10^{-30}$ & $1.38 \times 10^{-7}$ & $1.62
\times 10^{-4}$ & $2.24 \times 10^{-11}$ \\
2 & $5.56 \times 10^{-30}$ & $4.03 \times 10^{-7} $ & $3.58 \times
10^{-3}$ & $1.44 \times 10^{-9}$ \\
3 & $1.06 \times 10^{-29}$ & $7.69 \times 10^{-7}$ & 0.0115 &
$8.84 \times 10^{-9}$ \\
4 & $1.63 \times 10^{-29}$ & $1.18 \times 10^{-6}$ & 0.022 & $2.60
\times 10^{-8}$ \\
5 & $2.20 \times 10^{-29}$ & $1.59 \times 10^{-6}$ & 0.031 & $4.93
\times 10^{-8}$ \\
$\infty$ & $1.36 \times 10^{-28}$ & $9.82 \times 10^{-6}$ & 0.116
& $1.14 \times 10^{-6}$\\
\end{tabular}
\end{ruledtabular}
\end{table*}

\begin{table*}
\caption{\label{table2} Same as Table~\ref{table1} for $^{40}$K.}
\begin{ruledtabular}
\begin{tabular}{ccccc}
$p=l=q$ & $E_F$(J) & $T_F$(K) & $T_C(0)/T_F$ & $T_C(0)$(K) \\
\hline 1 & $1.71 \times 10^{-30}$ & $1.24 \times 10^{-7}$ & $3.93
\times 10^{-4}$ & $4.87 \times 10^{-11}$ \\
2 & $5.76 \times 10^{-30}$ & $4.17 \times 10^{-7} $ & $7.8 \times
10^{-3}$ & $3.26 \times 10^{-9}$ \\
3 & $1.19 \times 10^{-29}$ & $8.62 \times 10^{-7}$ & $0.023$ &
$1.98 \times 10^{-8}$ \\
4 & $1.93 \times 10^{-29}$ & $1.40 \times 10^{-6}$ & $0.04$ &
$5.54
\times 10^{-8}$ \\
5 & $2.72 \times 10^{-29}$ & $1.97 \times 10^{-6}$ & $0.054$ &
$1.06
\times 10^{-7}$ \\
$\infty$ & $2.12 \times 10^{-28}$ & $1.54 \times 10^{-5}$ & 0.155
& $2.37 \times 10^{-6}$\\
\end{tabular}
\end{ruledtabular}
\end{table*}

Tables~\ref{table1} and \ref{table2} illustrate the dramatic
effect that $p$ has on $T_F$ and $T_C(0)$. For both $^6$Li and
$^{40}$K, there is a two order of magnitude increase in the Fermi
temperature as $p$ is increased from $1$ to $\infty$. Furthermore,
in going from a harmonic trap ($p=2$) to a rigid box ($p=\infty$)
the Fermi temperatures increase by factors of 24 and 36 for $^6$Li
and $^{40}$K, respectively. Even more striking is the change in
$T_C(0)/T_F$, which increases by three orders of magnitude over
the full range of $p$. Altogether, this implies that by increasing
the confining power of the trap ($p$) one could, in principle,
increase $T_C(0)$ by as much as three orders of magnitude in
comparison to a harmonic potential. Note that in our calculation,
we take $E_F$ as the Fermi energy of an ideal Fermi gas. Including
the attractive atom-atom interaction will decrease $E_F$. However,
one must now include the Hartree-Fock term, $gn(0)$, in Eq.
(\ref{TCtrap}), which raises the critical temperature. We checked
this effect numerically and find that for the parameters used in
the paper, including the atom-atom interactions increases $T_C(0)$
by a factor of 2 to 3.

Physically, the increase in $T_F$ and $T_C(0)$ with increasing
confining power is a result of the trap being able to confine the
atoms to a smaller total volume, which thereby increases the local
density. For a rigid box, the atoms are confined to a volume of
$8abc$ regardless of the value of $E_F$. For a harmonic trap, the
total volume occupied by the gas corresponds to the extent of the
wave function for an atom in the highest occupied state with
energy $E_F\gg \hbar\omega$, which has a volume much larger than
the volume of the ground state wave function given by $\sim abc$.
In general, the total volume is given by $V(E_F)$. For the
isotropic form of the power law potential, $U({\bf
r})=\varepsilon(r/a)^p$, the volume scales as $V(E_F)\sim
N_F^{2/(p+2)}$. This explains the increase in $T_F$ and $T_C$
since, in the limit of a homogenous gas, they depend only on the
density of the fermions.

Alternatively, the increase in $T_C$ can also be explained by
examining the density of states for an atom with {\em total}
energy between $\epsilon$ and $\epsilon +d\epsilon$
\cite{bagnato},
\begin{equation}
\rho(\epsilon)=\frac{1}{4\pi^2}\left(\frac{2m}{\hbar^2}
\right)^{3/2}\int_{V(\epsilon)} \sqrt{\epsilon-U({\bf r})}\,d^3r.
\label{densitystates}
\end{equation}
Increasing the confining power of the trap reduces the volume of
phase space available to an atom with energy $\epsilon$,
$V(\epsilon)$, and as a result, the number of states is reduced.
In fact, it is easy to show that $\rho(\epsilon)\sim
\epsilon^{\delta -1}$. The reduction in the density of states
causes $E_F$ to be increased since the fermions are forced to
occupy higher energy states in order to accommodate all $N_F$
atoms. In the Thomas-Fermi approximation, the density of states on
the {\em local} Fermi surface at the center of the trap is
$\rho_F(0)=mk_F(0)/2\pi^2\hbar^2 \sim \sqrt{E_F}$. Consequently,
an increase in the Fermi energy, $E_F$, increases the local
density of states and therefore $T_C(0)$, as can be seen from Eq.
(\ref{TCtrap}).

Both Eq. (\ref{TC}) and Eq. (\ref{number}) are valid in the limit
of a dilute gas, $|a|n(0)\ll 1$. This approximation begins to
break down for $p>5$ when $|a|n(0)\gtrsim 0.2$ for both cases
considered here. There are, however, two reasons why this need not
be of any great concern. First, the neglect of the Hartree-Fock
mean-field in Eq. (\ref{density}) underestimates the actual
density of the gas at the center of the trap because the
interactions  are attractive. The fact that attractive
interactions increase the local density for trapped gases is well
known for Bose-Einstein condensates \cite{dalfovo} and has been
previously noted for fermions \cite{Li6}. Secondly, Heiselberg has
shown that in the regime of intermediate densities, $|a|n>1$,
$T_C$ becomes a finite fraction of $T_F$ with values of
$T_C/T_F\gtrsim 0.1$ \cite{heiselberg2}. Therefore, for large
confining powers, our results underestimate the actual value of
$T_C(0)$.

Finally we remark that from an experimental point of view,
changing the confining power of the trapping potential appears to
be realistic. The generation of power law traps would most easily
be accomplished using far-detuned optical dipole traps, which have
the added benefit of producing a confining potential that is
independent of the hyperfine state of the atoms. In particular,
the generation of higher-order Bessel beams with radial intensity
profiles proportional to $J_l^2 (k_r r)$, where $J_l$ is a Bessel
function of integer order and $k_r$ is the radial component of the
wavevector, have recently been produced using an axicon
\cite{axicon}. Bessel beams with $l=1$ to $4$ and radii for the
hollow core of the beam on the order of tens of micrometers were
created.  For blue-detuned beams and small $k_r r$, the radial
optical dipole potential experienced by the atoms is proportional
to $(k_r r)^{2l}$. Two perpendicular intersecting Bessel beams
with $l>1$ could be used to create an anharmonic potential. It is
worth noting that evaporative cooling of a two-component gas of
$^6$Li to temperatures below $T_F$ in an optical trap has recently
been demonstrated experimentally \cite{thomas}.

In conclusion we have examined the effect that a power-law
trapping potential has on the BCS transition temperature. We have
shown that by increasing the confining power of the trap, one can
obtain values of $T_C$ that are several orders of magnitude larger
than the corresponding harmonic trap. The origin of the increase
is the ability of tighter traps to confine the atoms to a smaller
total volume.

This work is supported in part by the US Office of Naval Research
under Contract No. 14-91-J1205, by the National Science Foundation
under Grant No. PHY98-01099, by the US Army Research Office, by
NASA, and by the Joint Services Optics Program.


\begin{references}
\bibitem{BEC}  M. H. Anderson {\it et al.}, Science {\bf 269}, 198 (1995);
K. B. Davis {\it et al.}, Phys. Rev. Lett. {\bf 75}, 3969 (1995);
C. C. Bradley {\it et al.}, Phys. Rev. Lett. {\bf 75}, 1687
(1995).

\bibitem{bardeen}  J. Bardeen {\it et al.}, Phys. Rev. {\bf 108}, 1175 (1957).

\bibitem{Li6}  H. T. C. Stoof {\it et al.}, Phys. Rev. Lett {\bf 76}, 10
(1996); M. Houbiers {\it et al.}, Phys. Rev. A {\bf 56}, 4864
(1997).

\bibitem{burnett} G. Bruun {\it et al.}, Eur. Phys. J. D {\bf 7},
433 (1999).

\bibitem{you} L. You and M. Marinescu, Phys. Rev. A {\bf 60}, 2324
(1999).

\bibitem{baranov} M. A. Baranov, JETP Lett. {\bf 64}, 301 (1996).

\bibitem{efremov} D. V. Efremov and L. Viverit, cond-mat/0108045.

\bibitem{heiselberg} H. Heiselberg {\it et al.}, Phys. Rev. Lett.
{\bf 85}, 2418 (2000).

\bibitem{bohn}  J. L. Bohn, Phys. Rev. A {\bf 61}, 053409 (2000).

\bibitem{holland}  M. Holland {\it et al.}, Phys. Rev. Lett. {\bf 87},
120406 (2001).

\bibitem{mackie}  M. Mackie {\it et al.}, Opt. Express {\bf 8}, 118
(2000); M. Mackie {\em et al.}, e-print physics/0104043.

\bibitem{truscott}  A. G. Truscott {\it et al.}, Science {\bf 291}, 2570
(2001); F. Schreck {\it et al.}, Phys. Rev. Lett. {\bf 87}, 080403
(2001).

\bibitem{thomas} S. R. Granade {\it et al.}, cond-mat/0111344.

\bibitem{jin}  B. DeMarco and D. S. Jin, Science {\bf 285,} 1703 (1999).

\bibitem{jin2} B. DeMarco {\it et al.}, Phys. Rev. Lett. {\bf 86}, 5409
(2001).

\bibitem{evaporate} M. J. Holland {\it et al.}, Phys. Rev. A {\bf
61}, 053610 (2000).

\bibitem{bagnato} V. Bagnato {\it et al.}, Phys. Rev. A {\bf 35}, 4354
(1987).

\bibitem{gorkov} L. P. Gorkov and T. K. Melik-Barkhudarov, Sov.
Phys. JETP {\bf 13} 1018 (1961).

\bibitem{lifshitz} E. M. Lifshitz and L. P. Pitaevskii,
{\it Statistical Physics Pt. 2 } (Pergamon, Oxford, 1980).

\bibitem{butts} D. A. Butts and D. S. Rokshar, Phys. Rev. A {\bf
55}, 4346 (1997).

\bibitem{mli} M. Li {\it et al.}, Phys. Rev. A {\bf 58}, 1445 (1998).

\bibitem{dalfovo} F. Dalfovo {\it et al.}, Rev. Mod. Phys. {\bf
71}, 463 (1999).

\bibitem{heiselberg2} H. Heiselberg, Phys. Rev. A {\bf 63}, 043606
(2001).

\bibitem{axicon} J. Arlt and K. Dholakia, Opt. Comm. {\bf 177}, 297
(2000).

\end{references}
\end{document}